# Spin dissymmetry in optical cavities

**ABSTRACT**. We introduce the spin dissymmetry factor, a measure of the spin-selectivity in the optical transition rate of quantum particles. This spin dissymmetry factor is valid locally, including at material interfaces and within optical cavities. We design and numerically demonstrate a metasurface optical cavity with three-fold rotational symmetry that maximizes spin dissymmetry, thereby maximizing the spin-selective radiative coupling of a cavity-coupled emitter. We also show the near-field and far-field response of spin and chiral dipoles to these cavities that preferentially enhance either spin or chirality. Our approach emphasizes the difference between spin and chirality in the near-field and reveals a compact parameter for designing more efficient quantum optical devices.

Priyanuj Bordoloi[1†*], Jefferson Dixon[2,3†*], Zachary N. Mauri[1], Christopher J. Ciccarino[1], Feng Pan[1], Tony Low[4], Felipe H. da Jornada[1], Jennifer A. Dionne[1*].

[1] Department of Materials Science and Engineering, Stanford University, Stanford, California 94305, USA

[2] Department of Mechanical Engineering, Stanford University, Stanford, California 94305, USA

[3] Johns Hopkins University Applied Physics Laboratory, Laurel, Maryland 20723, USA

[4] Department of Electrical and Computer Engineering, University of Minnesota, Minnesota 55455, USA

† Contributed equally

*Corresponding authors: Priyanuj Bordoloi, Jefferson Dixon, Jennifer Dionne.

**Email:** pbord@stanford.edu, jeff.dixon@jhuapl.edu, jdionne@stanford.edu

## I. INTRODUCTION

Circularly polarized light (CPL) plays a central role in light-matter interactions involving both spin and chirality. In quantum optical systems, CPL provides a direct basis for spin-selective excitation and emission, making it central to the optical preparation, control, and readout of spin-selective states, including in optically addressable spin qubits [1–5]. In chemistry, CPL acts as the primary probe of molecular chirality because of its selective interaction with the mirror symmetry of matter, a defining feature of stereochemistry and molecular function [6]. This shared relevance makes CPL a common tool used in both spin and chiral systems, and in the plane-wave limit the two can often be used interchangeably.

The apparent equivalence between CPL in spin and chiral systems, however, is incomplete. In the near-field of nanophotonic structures, where the wavevector is not uniformly defined, the usual identification of CPL with optical chirality becomes ambiguous [7,8]. Local formulations derived from Maxwell fields have clarified that optical spin and chirality are closely related but symmetry-distinct quantities [9,10]. In particular, the local chiral response of the field is captured by Kuhn's dissymmetry factor, which has been foundational to decades of development of chiral optical cavities, molecular detection, and enantioselective synthesis [11–18]. Yet because Kuhn's dissymmetry factor characterizes the chiral response of the field without independently resolving its spin character, it cannot fully capture interactions in which spin and chirality play distinct roles, motivating the need for a complementary local quantity defined in a spin basis.

Here we present the spin dissymmetry factor, which is a normalized local measure of spin-resolved optical field density. Derived from time-dependent perturbation theory, the resulting quantity is directly related to the transition rate of a particle along its spin quantization axis, with the magnitude and sign of the spin dissymmetry factor quantifying the local asymmetry between opposite circular transition channels. We show that the spin dissymmetry factor can be maximized in a metasurface cavity relying on quasi-bound states in the continuum with at least three-fold rotational symmetry as a consequence of the topological distribution of optical currents. Moreover, we also compare its interaction to an optical chirality-enhancing (Kerker) metasurface in the vicinity of a spin and chiral dipole.

*Contact authors: pbord@stanford.edu, jeff.dixon@jhuapl.edu, jdionne@stanford.edu

## II. HELICITY OPERATOR FORMALISM

The relationship between chirality and spin is defined in quantum field theory through the helicity operator,

$$\hat{h} = \widehat{\boldsymbol{S}} \cdot \frac{\boldsymbol{p}}{|\boldsymbol{p}|}, \tag{1}$$

where the helicity operator, $\hat{h}$, is defined as the projection of the spin operator $\widehat{\boldsymbol{S}}$ along the direction of the particle momentum $\boldsymbol{p}$ [19,20]. Bold letters indicate vectors, so that helicity $\hat{h}$ is a pseudoscalar arising from the three-dimensional space $\boldsymbol{S} \otimes \boldsymbol{p}$ when spin is transverse to the direction of linear momentum. For massless particles such as photons, optical chirality is the directly observable field quantity arising from this helicity [21]. The parity-odd symmetry of chirality and parity-even symmetry of spin are presented in Table 1; the operators, associated eigenmodes, and interaction polarizabilities follow this same symmetry.

Consider a state vector composed of complex (time-varying) electric and magnetic fields that vary in space $r$ such that,

$$\boldsymbol{\Psi}(\mathbf{r}) = \frac{1}{2\sqrt{\omega}} \begin{pmatrix} \boldsymbol{E}(\boldsymbol{r}) \\ \boldsymbol{H}(\boldsymbol{r}) \end{pmatrix}. \tag{2}$$

Here, the units $c = \epsilon_0 = \mu_0 = 1$ are assumed, and the transversality condition of electric and magnetic fields is ensured by $\widehat{\boldsymbol{p}} \cdot \boldsymbol{\Psi}(\mathbf{r}) = 0$, with the momentum operator $\widehat{\boldsymbol{p}} = -i\nabla$ [22]. We can map the electric and magnetic fields from Eq. (2) in a spin or chiral basis by operating on them with Eq. (1). The time-averaged local expectation values of spin and chirality are then, respectively,

$$\boldsymbol{S} = \langle\boldsymbol{\Psi}|\hat{\mathrm{S}}|\boldsymbol{\Psi}\rangle = \frac{1}{4\omega} Im(\boldsymbol{E}^* \times \boldsymbol{E} + \boldsymbol{H}^* \times \boldsymbol{H}), \tag{3}$$

$$C = \langle\boldsymbol{\Psi}|\hat{h}|\boldsymbol{\Psi}\rangle = -\frac{1}{2\omega} Im(\boldsymbol{E}^* \cdot \boldsymbol{H}), \tag{4}$$

where $\boldsymbol{S}$ and $C$ vary spatially as a function of $r$, and the asterisks indicate the complex conjugate. Spin density $\boldsymbol{S}$ can be defined for a two-dimensional plane, whereas chiral density $C$ can only be defined for a three-dimensional volume. For completeness, we also evaluate the local electromagnetic energy density,

$$W = \langle\boldsymbol{\Psi}|\omega|\boldsymbol{\Psi}\rangle = \frac{1}{4}(|\boldsymbol{E}|^2 + |\boldsymbol{H}|^2). \tag{5}$$

The values for spin density $\boldsymbol{S}$, chiral density $C$, and electromagnetic energy density $W$ are valid locally even when the wavevector is not well defined, such as in the near-field of subwavelength cavities (see Section I in [50]).

| | P-T Symmetry | Local Density | Enhancement Factor |
|---|---|---|---|
| Purcell Factor | + : + | $W = \dfrac{1}{4}(\lvert\boldsymbol{E}\rvert^2 + \lvert\boldsymbol{H}\rvert^2)$ | $f \propto \dfrac{W_E}{W_{E0}}$ |
| Spin Dissymmetry Factor | + : - | $\boldsymbol{S} = \dfrac{1}{4\omega} Im(\boldsymbol{E}^* \times \boldsymbol{E} + \boldsymbol{H}^* \times \boldsymbol{H})$ | $s \propto \dfrac{S_E}{S_{E0}}$ |
| Kuhn Dissymmetry Factor | - : + | $C = -\dfrac{1}{2\omega} Im(\boldsymbol{E}^* \cdot \boldsymbol{H})$ | $g \propto \dfrac{C_{EH}}{C_{EH0}}$ |

**Table 1.** Symmetry comparison of optical cavity enhancement factors. The "0" subscript indicates that the value is taken in free-space for normalization.
Parity (P) and Time-reversal (T) symmetry transformations are indicated as even (+) or odd (-).

## III. SPIN SELECTION RULES

When electromagnetic fields are confined in an optical cavity, the local density of optical states (LDOS) is modified. The transition rate of particles excited by light (*i.e.*, rate of optical absorption) is proportional to this local density of optical states. This is known as Fermi's Golden Rule, written for an interaction

Hamiltonian in the Coulomb gauge ($\nabla \cdot \boldsymbol{A} = 0$) with the form [23,24],

$$\hat{H}^{int} = \frac{e}{m_e c}\boldsymbol{A}^{\pm}(\boldsymbol{r},t)\cdot\boldsymbol{p}, \tag{6}$$

where $e$ is the charge of an electron, $m_e$ is the mass of an electron, $p$ is the momentum vector, and we have omitted the diamagnetic term proportional to $A^2$. The vector potential in this choice of gauge is,

$$\boldsymbol{A}^{\pm}(\boldsymbol{r},t) = A_0\hat{\epsilon}^{\pm}e^{i(\boldsymbol{q}\cdot\boldsymbol{r}-\omega t)}\,. \tag{7}$$

Here, the spatial dependence of the external field is related to the wavevector of the incident light, and the polarization vector $\hat{\epsilon}^{\pm}$ determines the handedness of the photon. By restricting our focus to near-field interactions, we consider only the vector potential at a point $r_0$, creating a local definition. This simplification yields a result equivalent to that which is obtained from the long wavelength approximation ($\boldsymbol{q} \approx 0$).

Treating the light-matter interaction as a perturbation to a ground-state many-body Hamiltonian, we then apply time-dependent perturbation theory to arrive at [19],

$$\Gamma^{\pm}_{v\to c}(k) = \frac{2\pi e^2 A_0^2}{\hbar c^2 m^2}\left|\langle f\boldsymbol{k}|\hat{\epsilon}^{\pm}\cdot\boldsymbol{p}|i\boldsymbol{k}\rangle\right|^2$$

$$\times\,\delta\big(E_c(\boldsymbol{k}) - E_v(\boldsymbol{k}) - \hbar\omega\big)\big(f_v(\boldsymbol{k}) - f_c(\boldsymbol{k})\big), \tag{8}$$

for the transition rate of an electron from the valence band $v$ to the conduction band $c$. We can then obtain a measure of the spin-selectivity of such excitations at each point in the dispersion,

$$\eta_{v\to c}(\boldsymbol{k}) = \frac{Im[\boldsymbol{p}^*_{cv}(\boldsymbol{k})\times\boldsymbol{p}_{cv}(\boldsymbol{k})]}{|\boldsymbol{p}_{cv}(\boldsymbol{k})|^2}, \tag{9}$$

which uses the momentum matrix element of the transition, $\boldsymbol{p}_{cv}(k)$, promoting the electron from the valence band to the conduction band. This process underlies the origin of spin-selection rules in a variety of quantum materials, most notably the exciton transitions of monolayer transition metal dichalcogenides (2D TMDCs). More generally, this describes the relative absorption of light into one eigenstate of a Kramers' degenerate pair when time-reversal symmetry is broken. Finally, we maximize the light-matter interaction Hamiltonian (Eq. 6) dot product by aligning the vector potential (Eq. 7) of a nonmagnetic transition with the momentum vector of the transition (Eq. 9) using the electric-dipole approximation. This is solved using classical electric fields of the form,

$$\boldsymbol{s}(\boldsymbol{r}) = \frac{Im(\boldsymbol{E}(\boldsymbol{r})^*\times\boldsymbol{E}(\boldsymbol{r}))}{|\boldsymbol{E}(\boldsymbol{r})_0|^2}, \tag{10}$$

and projecting onto the spin (quantization) axis,

$$s = \frac{Im(E_{\parallel}^*\cdot E_{\perp} - E_{\perp}^*\cdot E_{\parallel})}{|E_0|^2}. \tag{11}$$

Eq. (11) maximizes optical interaction with a general spin-selective transition in matter, which we refer to as the (local) *spin dissymmetry factor* (spin enhancement factor in Table 1). This is the fundamental result of this report. The spin dissymmetry factor quantity spans [-1, 1] for monochromatic plane waves, where it is 1, -1 for circularly polarized light, and 0 for linearly polarized light. For optical cavities that increase the local density of optical states beyond that of free space, the absolute value exceeds 1.

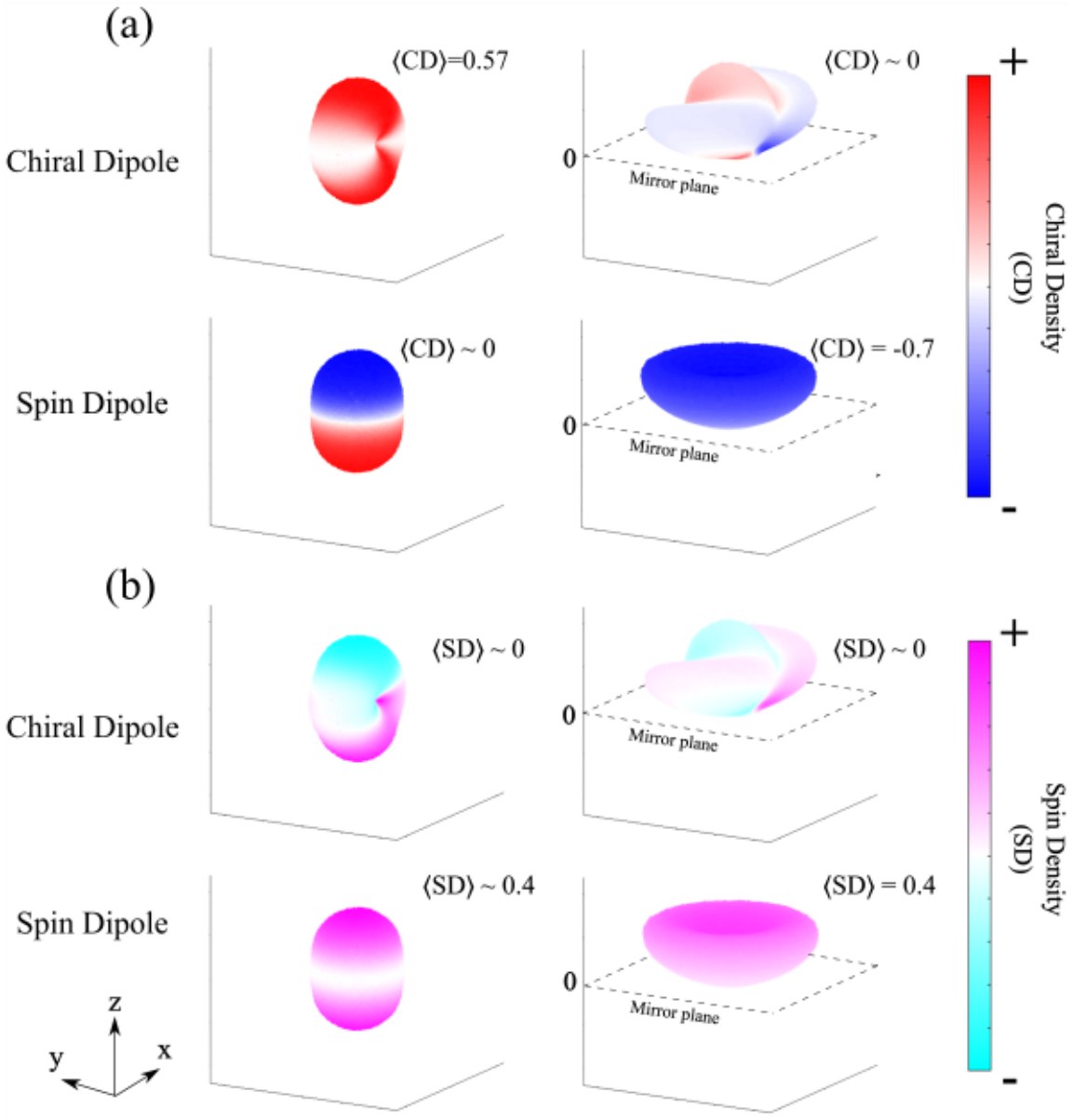

FIG 1. (a) Chiral density and (b) Spin density radiation profiles of chiral and spin dipoles with and without a mirror plane placed λ/2 below, where λ refers to the wavelength associated with the oscillation frequency of the dipoles. For details regarding the simulation set-up, refer to Appendix.

Like the Kuhn (chiral) dissymmetry factor (which is the normalized chiral energy density), and the Purcell factor (which is the normalized electric energy density), the spin dissymmetry factor has dimensionless units. This factor can be viewed as the normalized spin energy density or electric spin angular momentum, and it is compared directly to the Kuhn dissymmetry factor and Purcell factor in Table 1. Importantly, these quantities arise from the same dipole-level Fermi's-golden-rule structure and bear analogous relationships to cavity local density of states (LDOS), but they also correspond to different projections and inequivalent normalizations of light–matter coupling, such that appropriate re-normalization is necessary outside of the dipolar approximation (this often requires more complete knowledge of the density matrix). In this normalized perspective, we see that the primary difference between the spin dissymmetry factor and the Kuhn dissymmetry factor is in the separability of their electric and magnetic parts; spin can be viewed in a purely electric basis, whereas chirality is always electromagnetic. Spin objects (nonmagnetic) are modeled with the usual electric polarizability, while chiral objects are modeled with a (bi)anisotropic polarizability. In other words, the spin dissymmetry factor should be applied when interactions with matter preserve mirror symmetry, while the Kuhn dissymmetry factor should be applied when the interacting matter has broken mirror symmetry. For generalized spin and chiral constitutive relations, including distinguishing intrinsic/extrinsic chirality, planar or 2D chirality, and omega-type bianisotropy, see Section II of [50].

We visualize the fundamental symmetries of spin and chiral light-matter interactions using a dipolar polarizability model. As one example, the spin dipole can be considered representative of a spin-valley coupled exciton transition in a monolayer TMDC, and the chiral dipole can be considered as any small (subwavelength) polarizable chiral molecule. Here, we simulate the chiral and spin dipoles in free space and observe their respective chiral and spin densities under two conditions: no reflection and perfect mirror reflection (Fig. 1). For a chiral (spin) dipole, we observe uniform sign chiral (spin) density in free space but nullified spin (chiral) density, consistent with their parity symmetry. Reflection destroys the chiral density of a chiral dipole, but for a spin dipole, the spin density is preserved. In other words, the chirality of light is flipped upon reflection, and the two counterpropagating waves interfere destructively. The spin of light is not flipped upon reflection. Interestingly, the chiral density of a spin dipole becomes uniform in sign upon reflection, which is consistent with observations of chirality-induced spin selectivity in physical chemistry [25]. We observe that the magnitude of spin and chiral densities is necessarily dependent on the symmetry of the interaction, emphasizing the role of considering the symmetries of a cavity system along with its near-field light confinement. We return to these dipolar polarizability models in Section V to study the same interaction symmetries using their metasurface cavity equivalents.

## IV. A METASURFACE WITH MAXIMIZED SPIN DISSYMMETRY

We explore a model system to illustrate spin dissymmetry in a subwavelength optical cavity. Our high-Q dielectric metasurface is studied as an optical cavity to enhance coupling with a quantum particle possessing out-of-plane spin using a FEM solver (COMSOL).

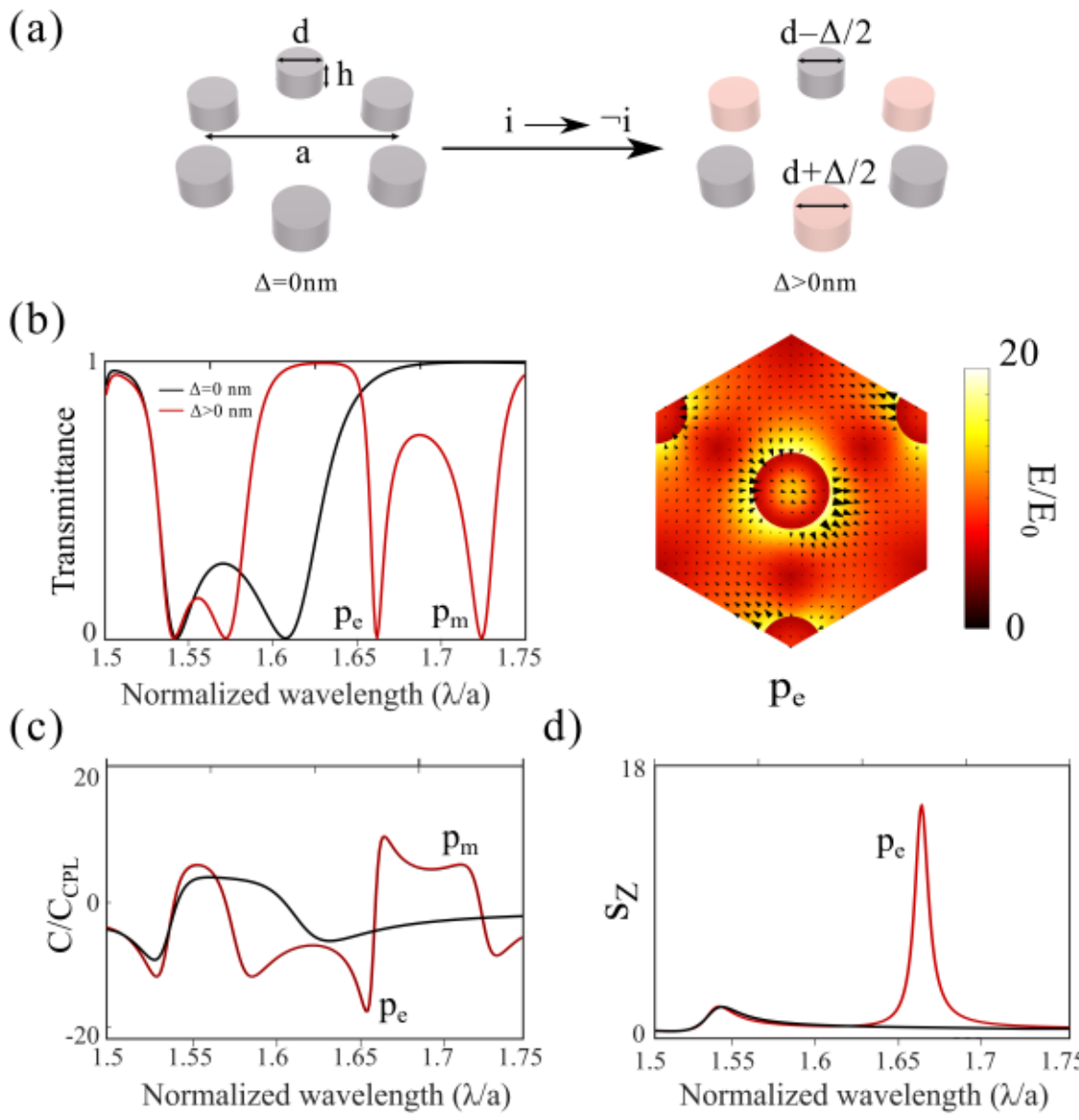

FIG 2. Illustration of honeycomb lattice metasurface and high-Q resonances. a) Schematic overview of honeycomb lattice metasurface with broken inversion symmetry, with period, a, and geometric asymmetry, $\Delta$. b) Transmittance spectra under circularly polarized light (CPL) illumination before (black) and after (red) inversion symmetry breaking. Right: electric field enhancement at the metasurface center plane on-resonance at $\lambda/a \sim 1.66$, with arrows indicating Poynting vector directions. (c) Average chiral density and (d) average spin density at the metasurface center plane as a function of wavelength.

Consider a dielectric metasurface composed of disks arranged in a staggered hexagonal (i.e., honeycomb) lattice, as shown in Fig. 2a. The dimensions of the metasurface are normalized by the lattice constant $a$, where the disk height $h = a/3$ and the disk diameter $d = 2a/9$; the diffraction limit is at $D = 3a/2$. We manipulate the inversion symmetry of the lattice by changing the diameter of neighboring disks, resulting in resonances in transmission under CPL illumination, as shown in Fig. 2b. When all the disks are the same diameter, we observe two broad resonances that correspond to the ordinary dipolar Mie modes. When neighboring disks are of different diameters (i.e., inversion symmetry is broken; $\Delta$>0), we observe two additional high-Q resonances. These two high-Q modes are symmetry-protected quasi-bound states in the continuum (q-BICs), and the linewidth of their resonances can be tuned using the difference in disk diameters; we show the electric field distribution at one of these high-Q modes in Fig. 2b as a representative example. We normalize the disk diameter into the asymmetry parameter $\delta = \Delta/d$ so that the q-BIC quality factors follow the scaling relation $Q \propto 1/\Delta^2$. Here, we choose a type of symmetry-breaking that preserves rotational symmetry, but it can be extended to any geometric perturbation within the unit cell, depending on the application [26–29].

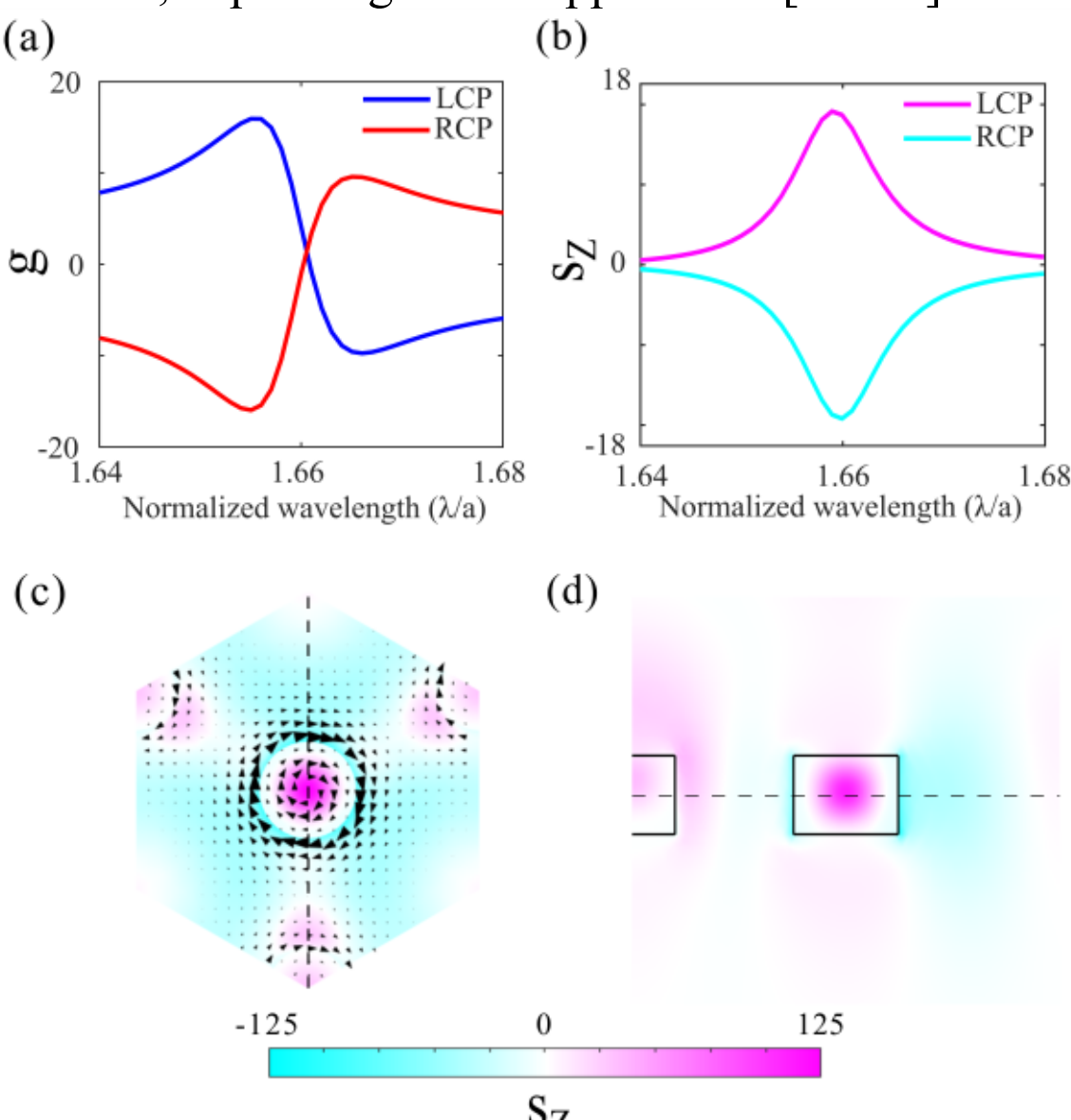

FIG 3. Chiral dissymmetry and spin dissymmetry of the simulated metasurface with LCP and RCP illumination. a) The chiral (Kuhn's) dissymmetry factor showing minimal chirality preservation on-resonance out-of-plane component, and b) The spin dissymmetry factor, showing maximal spin preservation on-resonance (electric-type mode) with both types of CPL illumination. c) The normalized dissymmetry factor at a plane through the center of the metasurface with arrows indicating the direction of the optical currents (Poynting vectors), taken on-resonance at $\lambda/a \sim 1.66$. d) The normalized dissymmetry factor evaluated in a plane perpendicular to the plane of panel (c) and passing through the dashed line, also taken on-resonance at $\lambda/a \sim 1.66$.

We plot the normalized spin dissymmetry and compare it to the normalized Kuhn (chiral) dissymmetry through the center plane of the metasurface in Fig. 2c,d. We observe uniform-sign enhancement of normalized spin dissymmetry under CPL illumination, whereas Kuhn (chiral) dissymmetry is both positive and negative across the q-BIC resonance. Dielectric Mie particles, like the disks in our metasurface, exhibit both electric-type and magnetic-type eigenmodes; the electric-type mode that we consider here has in-plane electric field vectors spinning to generate single-sign out-of-plane spin as shown in Fig. 3c. The spin density only requires that the near-field preserves the circular rotation of CPL, which is achieved with a cavity possessing rotational symmetry $C_{n\geq3}$ [30]. Chiral density, however, requires co-occurrence of the electric and magnetic fields in addition to their rotation; the asymmetric Fano line shape reflects the interference between electric and magnetic oscillations across the resonance [31]. The requirement of $C_{n\geq3}$ rotational symmetry aligned with the axis of propagation is difficult to achieve in on-chip photonics due to the symmetry-breaking of the substrate, making photonic crystals and metasurfaces particularly attractive for spin and chiral optics.

In Fig. 3a,b, we show the averaged normalized dissymmetry factors at a plane through the center of the metasurface upon CPL illumination spectrally across the q-BIC mode shown in Fig. 2b. We observe the single sign enhancement for spin dissymmetry compared to chiral (Kuhn's) dissymmetry for both types of CPL illumination. In Fig. 3c, we plot the spin density and optical currents for the same mode. Closer inspection of the spatial distribution of optical currents reveals that rotationally symmetric q-BIC resonances concentrate light selectively in regions of spin with a single sign, thereby maximizing spin dissymmetry. Here, we see that the optical currents circulate about the centers of large spin dissymmetry and avoid centers of low spin dissymmetry. Similarly, the current centers that are located over regions of high (low) energy density correlate with centers of high (low) dissymmetry, and this persists throughout the mode volume as observed in the perpendicular cross-section in Fig. 3d. Ultimately, this behavior is a consequence of the preferential distribution of topological charge introduced by the rotationally symmetric q-BIC mode, and we expect this behavior in a variety of high-symmetry lattice configurations.

## V. COMPARING OPTICAL CAVITIES FOR SPIN VERSUS CHIRALITY

Extending our investigation of chiral and spin dipoles suspended in free-space, we further explore their interaction with a Spin Metasurface and Kerker Metasurface (SM & KM). While both metasurfaces are designed to preserve rotational symmetry, the Spin Metasurface is designed for perfect reflection, and the Kerker Metasurface is designed for perfect transmission. These conditions match the theoretical boundaries that were explored in Fig. 1. For the Spin

Metasurface, we use the same design described in Section IV, and show its spin dissymmetry and near-field electric field density in the first column of Fig. 4a. For the Kerker Metasurface, we adjust the geometric features of the dielectric disks, while maintaining the radial perturbation, to spectrally overlap electric- and magnetic-type Mie modes from Fig 2b, approximately satisfying the requirements of the first Kerker condition (also known as a Huygens metasurface), and resulting in uniform transmittance upon illumination [32–35]. For more details on the design and characterization of the metasurface, refer to Section VI in Supplemental Material [50]. In both cases, the normalized resonance wavelength is designed to be the same, $\lambda/a \sim 1.66$.

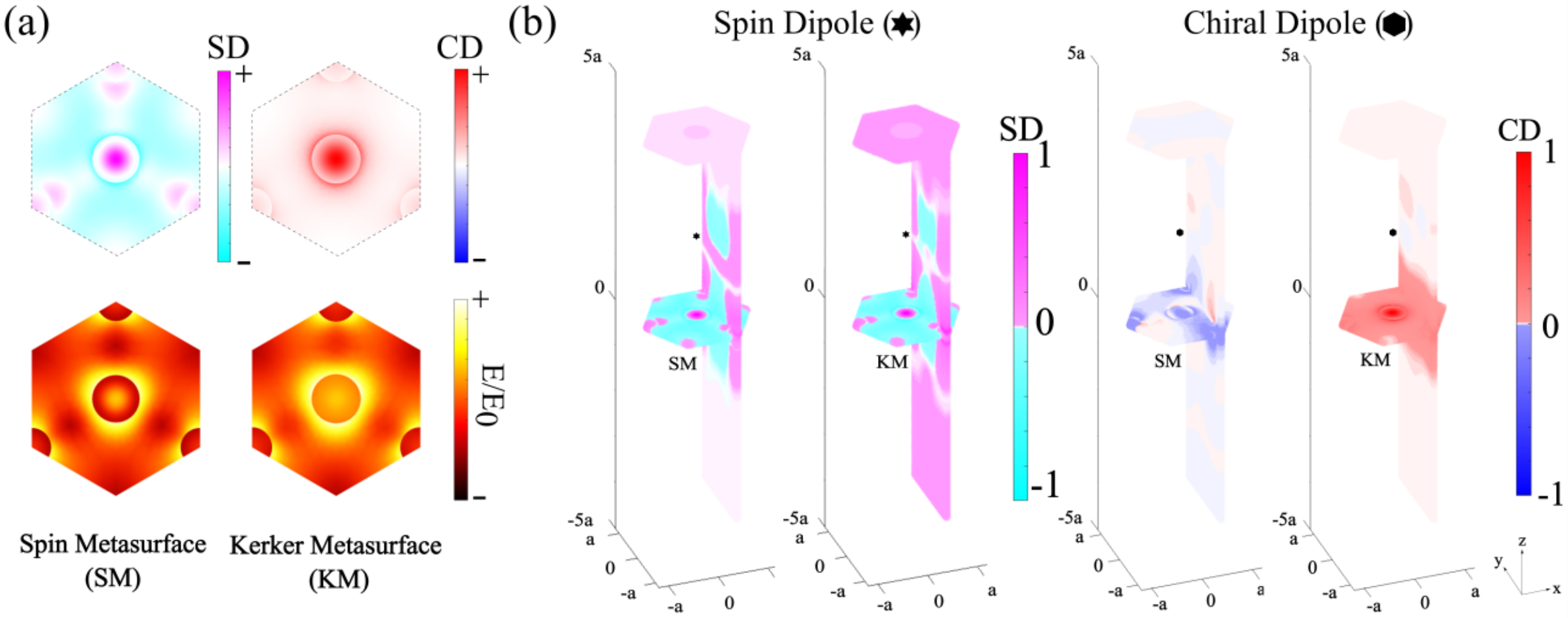

FIG 4. The near-field response of a spin and chiral dipole placed near a Spin Metasurface (SM) and a Kerker Metasurface (KM), taken at $\lambda/a \sim 1.66$ (on-resonance). (a) The near-field spin (chiral) density and local electric field density of a Spin (Kerker) Metasurface on resonance. (b) A chiral and spin dipole is placed λ above a SM and KM, and their response is observed through two X-Y planes (z = 0 & 4a) along with a Z-Y plane (x = a/2) through the local optical chiral and spin density.

We also explore how these near-field interactions are measured in the far-field in Fig 4b. A spin and chiral dipole are placed at a distance $\lambda$ above the metasurfaces, and we investigate the metasurface plane (z = 0), reflected plane (z = 4a), and the perpendicular Z-Y plane (x = a/2) by showing their respective spin and chiral density (SD & CD) after scattering off the metasurfaces. Notably, a chiral dipole does not get preferentially enhanced from the (reflective) SM due to the lack of a magnetic-type Mie mode on resonance, but it is uniformly enhanced upon scattering with a (transmissive) KM. The spin dipole, whether placed in a reflective cavity or transmissive cavity, maintains its spin density in the far-field and will result in circularly-polarized emission. In contrast, the chiral dipole only results in observable circularly-polarized emission if placed in a cavity that maintains its handedness, which in this case is achieved by the KM. The results follow the same trends observed in the pure dipolar responses from Fig. 1. This model shows that when a spin-selective emitter (such as a monolayer TMDC bright exciton) is placed in a cavity that enhances spin dissymmetry, the radiative coupling to the preferred circular transition channel is increased and the far-field circularly-polarized signal is maximized.

Likewise, when a chiral emitter (such as a fluorescently tagged chiral molecule) is placed in a cavity that enhances chiral dissymmetry, its far-field circularly-polarized signal is also maximized. Under absorption, both effects lead to circular dichroism. These observables may be confused for one another in simple 2-port measurement setups, or further convolved with one another under real-world conditions, such as with loss or a partially reflective substrate. We suggest that 4-port measurements can disentangle the origin of spin, chirality, and other anisotropies as described in the Supplemental Material [50]. The near-to-far-field projections we predict here are in agreement with experimental observations of cavity-mediated quantum emitters and chiral molecules. In quantum emitter cavities, the most efficient designs place the emitter in the center of a rotationally-symmetric distributed Bragg reflector stack, which provides electric field enhancement through reflection and preserves spin through rotational symmetry [36]. In chiral molecular detection, the circular dichroism sensitivity is maximized when the Kerker condition is met, preserving the handed relationship between electric and magnetic fields, and avoiding additional birefringence by again maintaining rotational symmetry [35]. We expect that more efficient interactions can be achieved by continuing to exploit these principles, for example, in bullseye configurations, open topological cavities, and multi-layer three-dimensional cavities. Similarly, cavities that, in themselves, break time-reversal or parity symmetry may selectively "lock" to a chosen spin or chirality, enhancing the sensitivity of circularly dichroic interactions [37–41].

## VI. CONCLUSIONS

We introduce the spin dissymmetry factor to quantify local enhancement (i.e., increase in the local density of optical states) in a spin basis, akin to how the Kuhn dissymmetry factor quantifies enhancement in a chiral basis. Like the Purcell factor, the spin dissymmetry factor can be enhanced by cavity confinement and resonant increases of the cavity quality factor [42]. Unlike the Purcell factor, this spin dissymmetry factor quantifies not only local field enhancement but also the direction of the local electric field rotation relevant to spin-selective optical transitions. In free space, circularly polarized light propagates with a defined direction of linear momentum, so there is typically no need to distinguish between its spin and chirality. Nanophotonics, however, requires local definitions where spin and chirality differ.

We design a high-Q optical cavity that distributes optical currents to maximize the spin dissymmetry, which minimizes the spin dephasing of a coupled particle. This advancement, especially in conjunction with the development of more robust quantum emitters (e.g., 2D TMDCs, Moiré superlattices, spin-entangled quantum dots), may help push quantum technologies that rely on single photons to more practical scales. In quantum computing, spin cavities may increase the determinism of single photon generation to reduce overall cryogenic requirements, increase coherence times for cluster states that massively reduce logical overhead, or serve as a substrate for nonlinear interactions that directly enact quantum logic [43]. Likewise, in quantum sensing, spin cavities may improve the signal amplitude and fidelity, especially at room temperature, or provide a higher-density platform to generate multiplexed quantum optical platforms. The target of high-density, high-fidelity, and controllable single photons is shared across quantum computing, sensing, and networking applications [44].

More broadly, the spin dissymmetry factor should generalize to any emitter whose optical interface is defined in a spin-selective circular basis. In that sense, beyond spin-valley-coupled excitons in TMDCs, the same figure of merit may provide a compact cavity-design rule for other spin-qubit platforms—including spin-selective quantum dots, trapped ions, defect centers, and related atom-like systems—whenever the relevant transition can be expressed through circular dipole channels along a chosen quantization axis. Under those conditions, maximizing spin dissymmetry should maximize the asymmetry of the local optical reservoir seen by the emitter and thereby improve spin-selective initialization, readout, and single-photon emission fidelity.

Here we highlight a useful distinction between local optical spin and optical chirality in nanophotonic cavities. Interactions governed by spin-selective circular electric-dipole coupling are naturally described by spin dissymmetry, whereas interactions governed by parity-sensitive electromagnetic handedness are naturally described by optical chirality. The design principles presented here may also be extended to other spin systems, such as topological objects like skyrmions and anyons, nonlinear and nonreciprocal systems, and quantum measurements [45,46]. Conveniently, these arguments can be described in semi-classical pictures with physical modes, providing a deeper intuition for understanding light-matter interactions.

## ACKNOWLEDGMENTS

The authors would like to thank Prof. David Barton III for the many stimulating conversations and insightful feedback on the topics of optical spin and chirality. The authors also acknowledge support from the Office of Naval Research under the Twist-Optics Multi-University Research Initiative (MURI) N00014-23-1-2567. P.B acknowledges support by the U.S. Department of Energy, Office of Basic Energy Sciences (DE-SC0021984).

## APPENDIX: SIMULATION DETAILS

The numerical calculations for the metasurface are performed using the Electromagnetics Module in COMSOL Multiphysics, which is a commercial Finite Element Analysis solver. For Fig. 2 and 3, the model is a 4-port simulation with periodic boundary conditions, where the ports assume Floquet periodicity. We illuminate the metasurface with circularly polarized light, either right-handed (RCP) or left-handed (LCP). The metasurface is comprised of disks with a refractive index of 3.5 and is lossless. Eq.

(10) and (11) also take the $E_0$ value for CPL in free space, such that the values for Eq. (10) and (11) are always 1 for CPL when there is no cavity (*i.e.,* $E_0 = E_{CPL,0}$).

We design a spin dipole by applying an electric point dipole with a dipole moment: (1/√2, i/√2, 0) [A.m]. A chiral dipole is designed by placing two electric dipoles (1/√2, 1/√2, 0) [A.m] & (1/√2, -1/√2, 0) [A.m] placed $\lambda$/4 distance apart. The interference of these two dipoles produces a chiral dipole. For the dipole simulations in Fig. 1, we use scattering boundary conditions placed several $\lambda$ away from the point source and PMLs to reduce any internal reflection. To simulate a perfect mirror, we place a perfect electric conductor (PEC) $\lambda$/2 distance away from the dipoles. For the dipole-metasurface simulations in Fig. 4, we have scattering boundary conditions on the boundary planes above and below the domain, and continuity conditions for the periodic boundaries around the metasurface. Similar studies have been done using these point dipoles and their radiation patterns [47–49].

## Supplemental Material:

# Spin dissymmetry in optical cavities

Priyanuj Bordoloi[1†*], Jefferson Dixon[2,3†*], Zachary N. Mauri[1], Christopher J. Ciccarino[1], Feng Pan[1], Tony Low[4], Felipe H. da Jornada[1], Jennifer A. Dionne[1*].

[1] Department of Materials Science and Engineering, Stanford University, Stanford, California 94305, USA

[2] Department of Mechanical Engineering, Stanford University, Stanford, California 94305, USA

[3] Johns Hopkins University Applied Physics Laboratory, Laurel, Maryland 20723, USA

[4] Department of Electrical and Computer Engineering, University of Minnesota, Minnesota 55455, USA

† Contributed equally

*Corresponding authors: Priyanuj Bordoloi, Jefferson Dixon, Jennifer Dionne.

## S1. Poynting vector and wavevector.

We use the same formalism as applied for Equations 3, 4 to arrive at the traditional Poynting vector and inspect its relationship to the local wavevector. By multiplying the energy, spin, and helicity operators simultaneously, we arrive at the Poynting vector,

$$\boldsymbol{P} = \langle \boldsymbol{\Psi} | \omega \hat{h} \hat{\boldsymbol{S}} | \boldsymbol{\Psi} \rangle = \frac{1}{2}\mathrm{Re}(\boldsymbol{E}^{*} \times \boldsymbol{H}) \tag{S1}$$

which differs from the local wavevector,

$$\boldsymbol{k} = \frac{\langle \boldsymbol{\Psi} | \hat{\boldsymbol{p}} | \boldsymbol{\Psi} \rangle}{|\Psi|^{2}} . \tag{S2}$$

***P*** and ***k*** will point in the same direction in vacuum, but in optical cavities or near material boundaries, the ***k*** may be complex, and need not point in the same direction as the power flow ***P***. Thus, in free-space where the wavevector is real, aligned with the direction of power flow, and homogeneous, the chirality and spin of light can be measured as the wavefront polarization transverse to the wavevector direction. Outside of these conditions, local (spatially varying) densities should be defined as real. Energy density (Equation 5), chiral density (Equation 4), and spin density (Equation 3) can be defined locally without an issue, because the final quantities are always real. Traditionally, the relationship between spin and chiral densities is derived by starting with the Poynting vector from Equation S1, which more explicitly shows the relationship between spin and orbital angular momentum rather than between spin and momentum [1]. Another consequence of starting from the Poynting vector is the arrival at a chiral flux as described in [2], where the chiral flux is the projection of the Poynting vector onto a spin plane, i.e., optical spin projected along power flow direction, and again is equivalent to the wavefront polarization [3].

## S2 Chiral and Spin Constitutive Relations

There is a common practice to define lower-dimensional forms of chirality (often labeled planar chirality, 2D chirality, or extrinsic chirality) that exhibit polarization selectivity in a circular basis (i.e., circular dichroism); however, these are more generally examples of mixed material anisotropy rather than chirality. This can be illustrated with the use of the constitutive relations, where the dyadic equation $[\boldsymbol{D}, \boldsymbol{B}]^{T} = A[\boldsymbol{E}, \boldsymbol{H}]^{T}$ includes the off-diagonal terms,

$$A = \begin{bmatrix} 0 & -is_z & -is_y & ic & -\omega_z & \omega_y \\ is_z & 0 & -is_x & i\omega_z & c & -i\omega_x \\ -is_y & is_x & 0 & -\omega_y & \omega_x & ic \\ -ic & \omega_z & -\omega_y & 0 & -is_z & -is_y \\ -\omega_z & -ic & \omega_x & is_z & 0 & -is_x \\ \omega_y & -\omega_x & -ic & -is_y & is_x & 0 \end{bmatrix} \tag{S3}$$

Where spin terms, $s_{x,y,z}$ , and omega-type bianisotropy terms, $\omega_{x,y,z}$ terms, x,y,z , are defined separably along each dimension, and the chiral term, c, is mapped inseparably across all dimensions simultaneously [4]. In other words, the scalar chiral term, c, couples electric and magnetic fields identically along all three axes, producing a pseudoscalar response that cannot be decomposed into independent directional components. The confusion arises when we consider mixtures of media, including periodic media such as metasurfaces, where otherwise achiral structures begin to exhibit polarization rotation or elliptical eigenmodes by mixing linear birefringence with omega-type bianisotropy [4]. Under realistic conditions, adding loss in these anisotropic structures can result in weak spin-selectivity, and adding a substrate to break the third mirror plane can result in weak chirality. A more rigorous test to elucidate the origin of circular dichroism is to perform at least a 4-port analysis, where transmission and reflection are measured in both forward and backward directions, or including measurements off-normal incidence (this is the test for "extrinsic chirality"). In the case of a more common 2-port measurement, this coupling matrix can be reduced into a two-dimensional subspace without removing the chiral terms, but even when chirality is measured in a subspace of the full coupling matrix, it still originates from a complete three-dimensional, isotropic property.

An interesting edge case arises in quasi-2.5-dimensional definitions of chirality, which can be useful in layered stacks of otherwise 2-dimensional media [5]. In these cases, chirality is still a 3-dimensional property, but compression along one dimension allows for certain 2-dimensional approximations – for example, considering that certain material properties are only continuous in-plane. Algebraically, it is tempting to map this coupling matrix onto a reduced (e.g., two-dimensional) subspace and refer to the remaining chiral terms as reduced-dimensionality chirality. However, these terms do not persist unless they also exist in all three dimensions. The out-of-plane coupling in some layered two-dimensional materials, if it breaks mirror symmetry, is enough to generate measurable optical chirality [6] .

**S3. Normalization under Fermi's Golden Rule.**

Our spin dissymmetry factor in Equation 11 is closely related to the spin angular momentum of light that is well-known in the literature on the spin-orbit interactions of light [7]. We extend the same formalism to the chiral density and electric energy density to emphasize their similarities. Following Fermi's Golden Rule, the normalization of the chiral density (Equation 4) is proportional to the Kuhn dissymmetry factor,

$$g \propto \frac{C}{|E_0|^2} \tag{S4}$$

And likewise, the normalization of the electric portion of the electromagnetic energy density (Equation 5) is proportional to the Purcell factor,

$$F_P \propto \frac{W_E}{|E_0|^2} \tag{S5}$$

Together, these factors describe Fermi's Golden Rule under three key scenarios: 1. parity-even and time-odd interactions (the spin dissymmetry factor), 2. parity-odd and time-even interactions (Kuhn's dissymmetry factor), and 3. parity-even and time-even interactions (the Purcell factor). The Purcell factor will often take a more detailed form, where the interaction cross-section between the electric field energy density and particle position, their spatial and spectral overlap, as well as nonradiative dissipation mechanisms, are considered. The spin dissymmetry factor can similarly be expanded, but their scaling with respect to the local density of optical states remains intact. In computation, the normalization in the denominator is equivalent to the energy density term in free space, i.e. $g \propto C/C_{CPL,0}$ .

**S4. Optical currents.**

In the cited work [1], the Poynting vector lines are referred to more generally as "optical currents," which we adopt in our description. Such vector contours are useful in describing closed surfaces, which is the case in vortices and rotational centers (known as C points). This convention also avoids the confusion of describing local power flow using the Poynting vector, when in fact the wavevector and power flow direction may not co-propagate.

**S5. Molecular polarizability.**

Consider the rate of absorption of a particle, $a = \omega\alpha''|E|^2$, where $\alpha''$ is the imaginary part of the particle polarizability. This solution splits degeneracy when time-reversal symmetry is broken, such that a spin-coupled particle experiences A+ and A- for absorption from left-CPL and right-CPL. The relative absorption of light into a single spin state (i.e., one eigenstate of the Kramers' degenerate pair) is then $\frac{2(a^+ - a^-)}{(a^+ + a^-)}$, which is equivalent to Equation 10 [2,7].

**S6. Kerker metasurface design**

To design the Kerker metasurface (KM), we change some of the dimensions of the spin metasurface (SM) to $r_{KM}$ = 0.23a, $h_{KM}$ = 0.293a, $a_{KM}$ = a (unchanged), $\delta_{KM}$ = 0.1 (unchanged). The resulting metasurface has the following transmission spectrum and surface average optical chirality through the center plane.

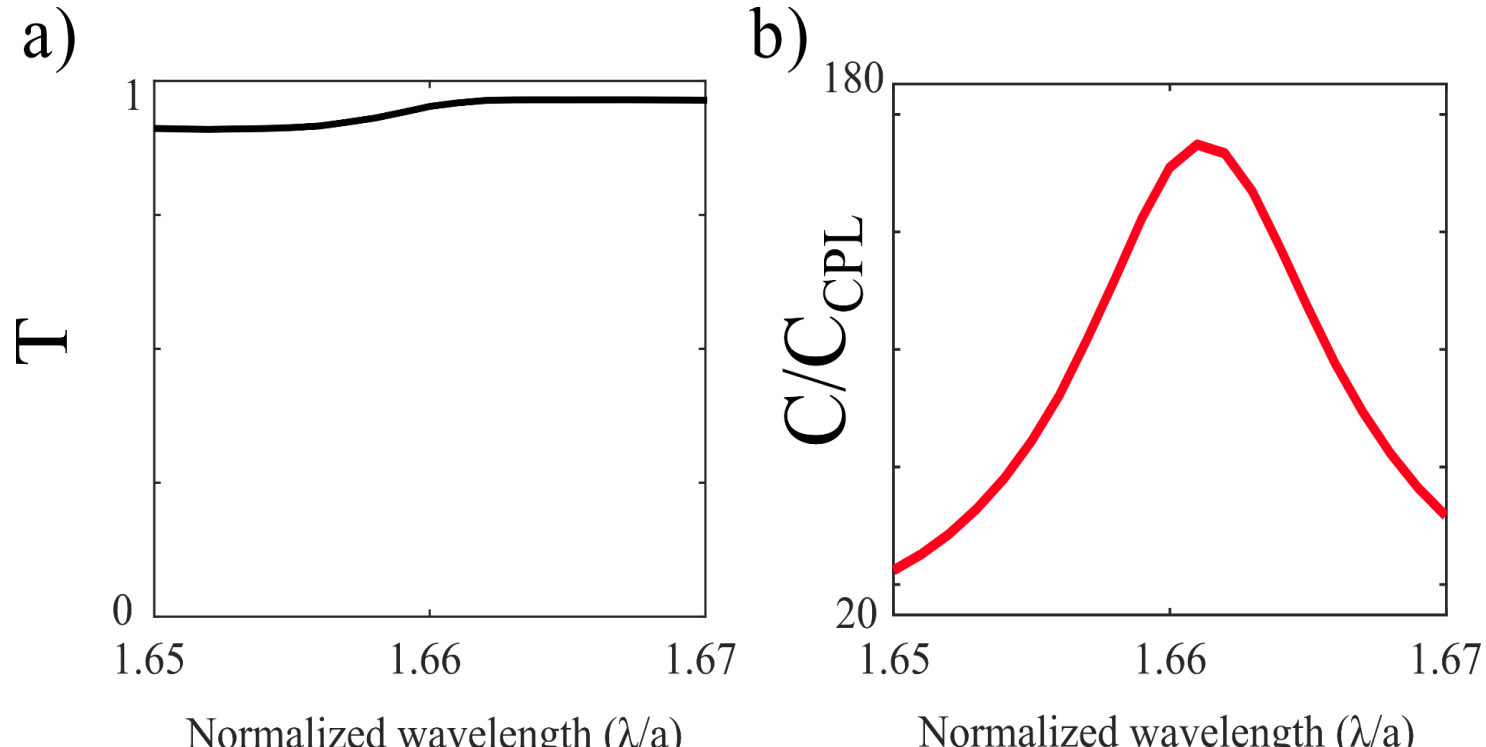

FIG S1: (a) Transmittance spectrum, and (b) surface average local density of optical chirality through the center plane of the metasurface.

This clearly shows that the designed metasurface satisfied the Kerker condition through uniform single-sign optical chirality enhancement.